\documentclass{article}
\usepackage{amsmath}
\usepackage{amsfonts}
\usepackage{amssymb}
\usepackage{graphicx}
\usepackage{tikz}
\usepackage{authblk}
\usepackage{hyperref} 
\usepackage[colorinlistoftodos,prependcaption,textsize=tiny]{todonotes}
\usepackage[linesnumbered,boxed]{algorithm2e}

\newcommand{\proof}{\noindent {\bf Proof}:\ }

\def\RR{\mathbb R}

\def\cH{\mathcal{H}}

\def\cI{\mathcal{I}}

\def\cD{\mathcal{D}}

\def\qed{{\hspace*{0mm}\hfill $\Box $\vspace{3mm}}}

\newtheorem{theorem}{Theorem}

\newtheorem{conjecture}[theorem]{Conjecture}
\newtheorem{corollary}[theorem]{Corollary}

\newtheorem{lemma}[theorem]{Lemma}

\newtheorem{remark}[theorem]{Remark}

\def\br{\color{red}\mathbf}

\def\br{\bar{r}}
\def\gs{\sigma}
\def\ts{\widetilde\sigma}
\def\tp{\tilde p}
\newcommand{\raf}[1]{(\ref{#1})}
\def\pinfty{\overline\infty}

\begin{document}
\author[1]{Endre Boros\thanks{\texttt{endre.boros@rutgers.edu}}}

\author[1,2]{Khaled Elbassioni\thanks{\texttt{khaled.elbassioni@ku.ac.ae}}}

\author[1,3]{Vladimir Gurvich\thanks{\texttt{vgurvich@hse.ru; vladimir.gurvich@gmail.com}}}

\author[3,4,5]{Mikhail Vyalyi\thanks{\texttt{vyalyi@gmail.com}}}

\affil[1]{MSIS \& RUTCOR, Business School, Rutgers University, 100 Rockafellar Road, Piscataway, 08854, NJ, USA} 

\affil[1,2]{Computer Science Department, Khalifa University of Science and Technology, 127788, Abu Dhabi, UAE }

\affil[3]{Higher School of Economics (HSE) of National Research University, Moscow, Russia} 

\affil[4]{Federal Research Center ``Computer Science and Control'' of the Russian Academy of Sciences}

\affil[5]{Department of Control and Applied Mathematics of Moscow Institute of Physics and Technology, Moscow, Russia} 
\date{}

\title{Two-person 
Positive Shortest Path Games 
Have Nash Equilibria in Pure Stationary Strategies}

\maketitle 

\begin{abstract} 
We prove that every finite two-person  
shortest path game, where the local cost of every move is positive for each player, has a Nash equilibrium 
(NE) in pure stationary strategies, 
which can be computed in polynomial time. 
We also extend the existence result to infinite graphs with finite out-degrees. 
Moreover, our proof gives that a terminal NE (in which the play is a path from the initial position to a terminal) exists 
provided at least one of the two players 
can guarantee reaching a terminal. 
If none of the players can do it, in other words, 
if each of the two players has a strategy that separates all terminals from the initial position $s$, 
then, obviously, a cyclic NE exists, 
although its cost is infinite for both players, 
since we restrict ourselves to 
positive games.
We conjecture that a terminal NE exists too,  
provided there exists a directed path from $s$ 
to a terminal. However, this is open.  

We extend our result to short paths interdiction games, where at each vertex, we allow one player to block some of the arcs and the other player to choose one of the non-blocked arcs. Assuming that blocking sets are chosen from an independence system given by an oracle, we give an algorithm for computing a NE in time $O(|E|(\log|V|+\tau))$, where $V$ is the set of vertices, $E$ is the set of arcs, and $\tau$ is the maximum time taken by the oracle on any input.

\end{abstract}   

\section{Introduction} 
\label{s0}
\subsection{Shortest Path Games and Nash Equilibria} \label{s-SP-Games}
Let $G = (V, E)$ 
be a finite directed graph (digraph) 
that may have loops, multiple arcs, and directed cycles. 
A fixed vertex  $s \in V$  is called {\em initial}. 
Furthermore, $V$ is partitioned into 
$3$  subsets 
$V = V_1 \cup V_2\cup V_T$ 
such that  $E$ contains no arc  $(u,v)$ leaving $u$ 
if and only if  $u \in V_T$. 
In the latter case $v$ is called {\em terminal} or 
a {\em terminal vertex}. 

For each $i \in \{1, 2\}$, 
a mapping  $r_i : E \rightarrow \RR$
assigns a real number $r_i(e)$  
to each arc $e \in E$,  called the $i$th local cost (or length) of $e$.
Given $i \in \{1,2\}$ and
a directed path $p$ from  $s$ 
to a terminal $v \in V_T$,  
we assign the real number 
$R_i(p) = \sum_{e \in p} r_i(e)$ 
and call it the $r_i$-effective cost (distance, or length) of $p$. 

The considered model can be naturally interpreted 
in terms of game theory. 
Digraph $G = (V=(V_1\cup V_2),E)$  models a two-person positional game $\Gamma=(G,s,V_T,r_1,r_2)$: 
its vertex $v \in V_i$  is a position controlled by player $i\in\{1,2\}$,  
and an arc $e = (u,v) \in E$  
is a legal move from $u$ to $v$. The game starts in a fixed (initial) position $s$; 
when the control is at vertex $u\in V_i$, player $i \in \{1,2\}$ chooses a move $(u,v) \in E$, 
after which the control moves to $v$. The game terminates when a vertex $t\in V_T$ is reached. 
The $i$th local cost $r_i(e)$ is interpreted as a cost of move $e$  for player $i \in \{1,2\}$. 

A play $p$ in this game is defined to be either a directed path from $s$ to some vertex  $t\in V_T$ (called $(s,t)$-path), or as an ``$s$-lasso", that is, a (possibly empty) directed path starting in $s$ and ending in a directed cycle. 
If the play is a path, then its {\em effective cost} $R_i(p)$ 
for each player $i \in [n]$ is assumed to be additive: 
$R_i(p) = \sum_{e \in p} r_i(e)$; if $p$ is an $s$-lasso, then $R_i(p)=+\infty$, $0$, or $-\infty$, depending on whether $R_i(p)>0$, $R_i(p)=0$ or $R_i(p)<0$, respectively. 
We assume that both players are {\it minimizers}, that is, 
they want to reduce their effective costs. 

We call such a game a two-person shortest path (SP) game. It is said to be {\em finite} if its digraph $G$ is finite 
and {\em positive} if its local costs are positive\footnote{Equivalently, all directed cycles 
have positive total costs for each player   \cite{Gal58}.}: 
$r_i(e) > 0$  for each player $i \in \{1,2\}$ and each move $e \in E$.

A strategy $\gs_i$  of a player $i \in \{1,2\}$  
is a mapping $\gs_i : V_i \rightarrow V$  
that assigns to each position $v \in V_i$ 
a move $(v, \gs_i(v)) \in E$  from $v$. We denote by $S_i$ the set of strategies of player $i\in \{1,2\}$. 
A pair $\gs = (\gs_1,\gs_2)$ 
is called a {\em strategy profile} or a {\em situation}. 
Every situation $\gs$ uniquely defines a play 
$p(\gs)$; 
each move $(u,v)$ of $p(\gs)$ is determined 
by the corresponding strategy $v =\gs_i(u)$  if $u \in V_i$.  

A situation $\gs = (\gs_1, \gs_2)$  is called 
a Nash Equilibrium (NE) if 
no  player $i \in \{1,2\}$ can improve his/her effective cost $R_i(p(\gs))$ 
by replacing his/her strategy $\gs_i$ by 
some other strategy $\gs'_i$, provided 
the  other player keeps his/her strategy unchanged, that is, for all $i\in\{1,2\}$,
$R_i(p(\gs) \leq R_i(p(\gs'))$, provided 
$\gs'$ may differ from  $\gs$ only in the $i$th entry. 
In other words, for each player $i \in \{1,2\}$ 
his/her strategy $\gs_i$  is the best response to the other player's strategy $\gs_{3-i}$. 
This means stability provided only one player 
(not a coalition) may try to deviate 
from the considered situation $\gs$. 
This is the simplest and most popular concept of solution 
in the non-cooperative game theory; 
it was introduced in 1950 by John Nash \cite{Nas50,Nas51}.

A NE $\gs$ in an SP game is called {\em terminal}  
if the corresponding play $p(\gs)$ 
is a path from $s$ to $V_T$ and 
{\em cyclic} if  $p(\gs)$  starts in $s$ 
and does not come to $V_T$. 
In this case it is an $s$-lasso 
that starts in $s$, proceeds with an initial path, which may be empty, and cycles as soon as it revisits a position. 
\footnote{In this case it will repeat the obtained 
directed cycle infinitely,  
because we restrict the players 
by their stationary strategies, which repeat 
a move whenever a position is revisited.} 

For a positive SP game, 
all cyclic plays are of infinite cost 
for each player.
We treat them as equivalent
and denote any of them by the same symbol $\pinfty$. 
Thus, every terminal play is cheaper 
than any cyclic one for each player. For $i\in\{1,2\}$, we say that player $i$ can cut $V_T$ from $s$, if he/she has a strategy $\gs_i$ such that the graph obtained from $G$ by fixing $\sigma_i$ (and deleting all arcs not selected by $\sigma_i$ at $v\in V_i$) does not contain a directed path from $s$ to any of the terminals (called $(s,V_T)$-path).
 
 The main result of this paper is the following.
\begin{theorem} 
\label{t-Bi-SP-game}
Every finite positive two-person SP game has a NE. Furthermore, such a NE can be computed in polynomial time. 
\end{theorem} 

Note, however, that this NE may be cyclic. 
Furthermore, a cyclic NE exists in a two-person SP game 
whenever each of the two players 
has a strategy cutting $V_T$  from  $s$. 
Indeed, if each player applies such a strategy 
then $\pinfty$ is the only 
possible outcome and we obtain an NE, 
although $\pinfty$ is of infinite cost for both players. 
We conjecture that a terminal NE exists too 
provided $G$  has a directed path from $s$ to $V_T$. 

\begin{conjecture} 
\label{cnj-Bi-SP-game}
Every finite positive two-person SP game has a terminal NE 
provided it has at least one terminal play.  
\end{conjecture} 



This conjecture
was proved for symmetric digraphs\footnote{$G$ is symmetric if  $(u,w) \in E$ 
whenever $(w,u) \in E$  and $u \not\in V_T$.} in \cite{BFGV23},
where it was also shown 
that requirement  $r_i(e) > 0$ 
is essential 
\cite[Figures 4 and 5]{BFGV23}. In this case the above conjecture 
can be extended and proven for an arbitrary number of players, not only $2$.
On the other hand, Theorem \ref{t-Bi-SP-game} cannot be generalized for 
the case of more than $2$ players \cite[Figure 2 and Table 3]{GO14}. 

\medskip

Finally, we observe that the existence result in Theorem~\ref{t-Bi-SP-game} can be extended to infinite graphs with finite out-degrees.
\begin{theorem} 
\label{t-Bi-SP-game-infinite}
Every positive two-person SP game where every vertex has finite out-degree has a NE. 
\end{theorem} 

Variants of shortest path games have been considered extensively in the literature, see, e.g., \cite{BGPW08,BT14,DPS17,P97,PB99,R13,TJABC21,VS02}, 
mostly focusing on pure history dependent strategies and 
solving by backward induction.

\subsection{Bi-shortest path Theorem and Conjecture}
The game-theoretic result stated above can also be interpreted purely in terms of graph theory. As before, let $G = (V, E)$ 
be a digraph, $s\in V$  be a given vertex, and $V_1\cup V_2\cup V_T$ be
a partition of $V$ into 
$3$  subsets 
such that all vertices in $V_T$ are terminals. For each $i \in \{1,2\}$, 
$r_i : E \rightarrow \RR$
be a mapping that assigns a local cost (or length) $r_i(e)$ to each arc $e \in E$, and for a directed path $p$ from  $s$ 
to a terminal $v \in V_T$,  
define the effective (or total) cost (or length) of $p$ to be 
$R_i(p) = \sum_{e \in p} r_i(e)$. Denote further by $\gs_i : V_i \rightarrow V$ a mapping that assigns to each vertex $u \in V_i$ a vertex  $v = \gs_i(u) \in V$  such that  $(u,v) \in E$  is an arc of $G$, 
and let $S_i$  be the set of such mappings. 

Assume that all local costs are positive 
($r_i(e) > 0$ for all $i \in \{1,2\}$  and $e \in E$). Fix a mapping $\gs_1 \in S_1$ and for each $v \in V_1$, 
choose the arc $(v,\gs_1(v)) \in E$ 
and delete from $G$ all other arcs 
$(v,w)$ going from $v$.  
Denote the obtained reduced digraph by $G(\gs_1)$ 
and denote by $P_2(\gs_1)$ the set of all 
{\em shortest paths} (SP) from $s$ to (some vertex in) $V_T$ 
with respect to the local lengths $r_2$,  
or in other words, minimizing the effective cost $R_2(p)$ over all $(s,V_T)$-paths $p$. 

It may happen that $\gs_1$ {\em cuts} 
$V_T$ from $s$, that is, digraph $G(\gs_1)$ 
has no $(s,V_T)$-paths. 
In this case we set $P_2(\gs_1) = \pinfty$, 
where $\pinfty$ is a special symbol, denoting a path of effective cost $+\infty$. 
Set $P_2 = \cup_{\gs_1 \in S_1} P_2(\gs_1)$. 
Then, by swapping the indices $1$ and $2$ define $P_1$ similarly.

It is easily seen that we can reformulate 
Theorem \ref{t-Bi-SP-game} and Conjecture  \ref{cnj-Bi-SP-game} 
in the graph theoretic terms as follows.
\begin{theorem} [Bi-SP Theorem]
\label{t-Bi-SP} 
$P_1 \cap P_2 \neq \emptyset$.
\end{theorem}

In other words, sets $P_1$ and $P_2$ 
always have a directed path or symbol $\pinfty$ in common.

Clearly Theorem~\ref{t-Bi-SP} follows immediately from Theorem~\ref{t-Bi-SP-game} if we recal the game-theoretic interpretation: if $p \in P_1 \cap P_2$ then  
there exists a situation $\gs = (\gs_1, \gs_2)$ 
such that $p = p_2(\gs_1) = p_1(\gs_2)$, where $p_{3-i}(\gs_i)\in P_{3-i}(\gs_i)$, for $i=1,2$. 
Hence, $\gs_1$ is a best response to $\gs_2$ and vice versa,  
that is, $\gs$ is a terminal NE. 

By definition, $\pinfty \in P_i$ for $i=1,2$ 
if and only if 
there exists a mapping $\gs_{3-i}$ 
cutting $V_T$ from $s$,  
that is, the reduced digraph in $G(\gs_{3-i})$ 
has no $(s,V_T)$-path. 
In particular, $\pinfty \in P_1 \cap P_2$ 
if and only if 
such cutting mappings $\gs_1$ and $\gs_2$ both exist. 
We conjecture that even in this case 
$P_1$ and $P_2$  have an $(s,V_T)$-path in common provided it exists in $G$.  

\begin{conjecture}[Bi-SP Conjecture~\cite{BFGV23,Gur21,GO14}] 
\label{cnj-Bi-SP} 
If $G$  has an $(s,V_T)$-path 
then sets $P_1$ and $P_2$ have such a path in common, that is, 
$(P_1 \cap P_2) \setminus \{\pinfty\} \neq \emptyset$. 
\end{conjecture}

Note that the case when 
$G$ has no $(s,V_T)$-path is trivial. 
Indeed, in this case 
all considered sets contain only 
$\pinfty$  
and no other elements.

As we know, 
$\pinfty \not\in P_i$  if and only if 
no  $\gs_{3-i} \in S_{3-i}$ cuts $V_T$ from $s$. 
If this happens for $i=1$ or $i=2$ 
then the Bi-SP Theorem and Conjecture coincide. 
Thus, the latter remains open only if 
there exist both $\gs_1 \in S_1$ and $\gs_2 \in S_2$ 
cutting  $V_T$ from $s$. 
Otherwise, the statement of Conjecture \ref{cnj-Bi-SP} 
follows from Theorem  \ref{t-Bi-SP}.

\subsection{Simplifying Assumptions} 
\label{ss-wlog}

Without loss of generality (wlog) we can assume the following. 

\begin{itemize}
\item [(i)] 
$|V_T|=1$. Indeed, merge all terminal vertices in $V_T$  
and replace them by a unique terminal $t$. 
After this, a directed path from $s$ to $V_T$ 
will be called an $(s,t)$-path.
\item [(ii)] 
An $(s,t)$-path exists.  
Indeed, otherwise 
$P_1=P_2=\{\pinfty\}$.
\item [(iii)]
Each vertex $v \in V$ and arc $e \in E$ is reachable from $s$. 
Indeed, otherwise it can be just deleted. 
\end{itemize} 

The next two simplifying assumptions  
also can be made without loss of generality  
and, although are not necessarily used in the proofs,  
they may help the reader to have a better understanding of the problem.

\begin{itemize} 
\item [(iv)]
All directed $(s,t)$-paths have pairwise distinct 
effective costs. 
Indeed, we can get rid of all ties by an 
arbitrary small perturbation of local costs.
After this, a shortest $(s,t)$-path 
is unique in $G(\gs_i)$, 
that is, $|P_i(\gs_{3-i})| = 1$ for $i = 1,2$. 
\item [(v)]
The induced subgraph $G[V \setminus \{t\}]$ 
is bipartite and  
for every arc $(u,w)$  in it, 
either $u \in V_1, w \in V_2$ or vice versa.
In other words, two players make moves alternately. 
Assume that $u,w \in V_1$ (resp., $u,w \in V_2)$. 
Then, subdivide this arc into two 
replacing  $(u,w)$  by  $(u,v)$ and $(v,w)$, 
where $v \in V_2$  (resp., $v \in V_1)$
and define positive local costs 
$r_k(u,v)$  and  $r_k(v,w)$  such that 
$r_k(u,v)+r_k(v,w)= r_k(u,w)$  for both $k \in \{1,2\}$. 
\end{itemize}

Obviously, (iv) and (v)  may only destroy a NE but cannot create one. 
Since we are studying conditions that imply 
the existence of an NE, we can apply  
(iv) and (v) wlog.

\subsection{Necessary Conditions for Theorem~\ref{t-Bi-SP-game-infinite} and Conjecture~~\ref{cnj-Bi-SP-game}} 
The two properties 
of an SP game required in Theorem~\ref{t-Bi-SP-game-infinite} (positivity and being defined on two players) are 
essential for existence of an NE.
A finite positive three-person 
NE-free SP game was constructed in 
\cite[Figure 2 and Table 3]{GO14}. 

A two-person finite NE-free SP game  
having some negative local costs 
was given in \cite{BFGV23}. 
Moreover, such an example was obtained 
also for an SP game with non-negative local costs, 
yet, this SP game has directed cycles 
whose all local costs for one of the two players are $0$ 
\cite[Figures 4 and 5]{BFGV23}.
\footnote{According to \cite{Gal58}, 
an $n$-person SP game can be replaced by a nonnegative one, 
by applying a potential transformation, 
whenever all its directed cycles 
have nonnegative total cost for all players.} 

Finally, note even 
a one-person positive but infinite SP game may be terminal NE-free (that is, Conjecture~\ref{cnj-Bi-SP-game} does not hold for infinite graphs). 
Let $G$ be a caterpillar,  
with local costs on the main path: 
$1, \frac{1}{4},\dots,\frac{1}{4^k},\dots$, and 
the local cost of the termination after $k=0,1,2,\ldots$ 
moves along the main path are 
$2, \frac{1}{2},\dots, \frac{2}{4^k},\dots$. 
The efficient cost of the play in $G$  
that terminates after $k$ steps is  
$1 + \frac{2}{3}(\frac12 + 4^{-k})$. 
It is monotone decreasing with $k$.  
Hence, the obtained one-person game 
has no shortest play and, thus, no terminal NE.
\footnote{Recall that 
wlog (assumption (i)) we can merge all terminal positions $V_T$
of the caterpillar $G$ 
getting a unique terminal $t$.}

\subsection{SP Games Are 
Stochastic Games 
with Perfect Information, Positive Local Costs,  
and 1-Total Effective Cost} 

The SP games can be viewed as 
a special case of the so-called 
deterministic stochastic games with perfect information 
introduced in 1957 by Gillette \cite{Gil57}.
In \cite{Gil57}, the {\em limit mean effective cost} was considered. 
Then {\em total effective cost} was introduced in \cite{TV98}; 
see also \cite{BEGM18} and \cite[Section 5]{GO14}.
Then, in \cite{BEGM17} a more general $k$-total 
effective cost was introduced: the limit mean and total 
effective costs correspond to $k = 0$ and  $k = 1$, respectively. 
In all cases the local costs are arbitrary 
(not necessarily positive) real numbers. 
The existence of a NE (a saddle point) 
for the two-person zero-sum case is a classical 
result of the stochastic game theory. 
It was shown for $k=0$ in \cite{Gil57,LL69}, 
in \cite{TV98,BEGM18} for $k=1$, and
in \cite{BEGM17} for arbitrary $k$. 
However, this result cannot be extended 
to the two-person non zero-sum case. 
An example for $k=0$ was given in \cite{Gur88}; 
then it was extended to all $k$ in \cite{BEGM17}. 

Interestingly, the SP games considered in the present paper  
correspond to the case $k=1$. 
Yet, we also assume that the local costs are positive.  
This assumption essentially simplifies general formulas 
for the effective 1-total cost 
(see \cite{BG03,BEGM17} and also \cite[Section 5]{GO14}) 
and, in accordance with Theorem \ref{t-Bi-SP-game} 
implies the existence of an NE 
in pure stationary strategies. 

\subsection{
Two-Person Short Paths Interdiction Games}

Given a digraph $G = (V, E)$, envisioned as a network, 
with a non-negative cost (length) function on its arcs
$r : E \rightarrow \RR_+$ 
and two terminals $s, t \in V$, 
an attacker may want to destroy all short directed paths from
$s$ to $t$ in $G$ by eliminating some arcs of $A$, 
given some limited resource that he can use
for such destruction.
This problem was considered 
in \cite{Gol78,HS21,IW02,KGZ06,KBBEGRZ08,MMG89,Phi93,Was95,Whi99,Woo93} 
under the name of 
{\it the short paths interdiction (or inhibition) problem}.   
It can be viewed as a zero-sum game. 
Here we extend it to the general two-person, 
not necessarily zero-sum, case. 

\medskip 

In particular, we consider a generalization of the two-person SP game introduced in Section~\ref{s-SP-Games}, where rather than having a partition of the vertex set $V$ into $V_1$ and $V_2$, we allow one player, say player $1$, to block some of the arcs at each vertex, and the other player to choose one of the non-blocked arcs. More formally, denote by $E(u)=\{(u,v)\in E~|~v\in V\}$ the set of arcs going out of vertex $u\in V$, and assume that at each vertex $u\in V\setminus\{t\}$, we are given an {\it independence (blocking) system}, that is, a subfamily of $E(u)$ closed under taking subsets. Let $\cI(u)\subseteq 2^{E(u)}$ be the family of {\it independent} sets of the blocking system at $u$, and 
denote by $\cD(u)$ the family of {\it dependent} sets (i.e., those that are not contained in any independent set in $\cI(u)$). 
A strategy of player $1$ (resp., player $2$) is a mapping $\sigma_1:V\to 2^E$ (resp., $\sigma_2:V\to 2^E$) which assigns to $u\in V\setminus\{t\}$ an independent set $\sigma_1(u)\in\cI(u)$ (resp., a dependent set $\sigma_2(u)\in\cD(u)$). We assume that $E(u)\not\in\cI(u)$, and hence $\cD(u)\ne\emptyset$, for all $u\in V\setminus\{t\}$. 
As before, given fixed start and terminal vertices $s$, $t$ and a pair of positive local cost functions $r_1,r_2:E\to\RR_+$, of players $1$ and $2$, respectively, the objective of player $i\in\{1,2\}$ is to ``agree'' with player $3-i$ on a path that minimizes the $r_i$-distance from $s$ to $t$ in the subgraph $G(\sigma_1,\sigma_2)=\big(V,\bigcup_{v\in V\setminus\{t\}}\sigma_2(v)\setminus\sigma_1(v)\big)$. More precisely, for $i\in\{1,2\}$, let $P_i(\sigma_1,\sigma_2)$ be the set of $r_i$-shortest paths in the graph $G(\sigma_1,\sigma_2)$. Then the cost of player $i$ corresponding to the situation $\sigma=(\sigma_1,\sigma_2)$ is defined to be $r_i(p)$ if there is a path $p\in P_1(\sigma)\cap P_2(\sigma)$ and $+\infty$, otherwise. The assumptions that $\sigma_1(v)\in\cI(v)$ and $\sigma_2(v)\in\cD(v)$, for all $v\in V$, imply that $\sigma_2(v)\setminus\sigma_1(v)\ne\emptyset$ and hence the game is well-defined. 
We call such a game $\Gamma=(G,s,t,\cI,r_1,r_2)$ an SP interdiction game. 
Note that we obtain an SP game as a special case if we set $\cI(u)=\{I\subseteq E(u):~|I|\le|E(u)|-1\}$ for $u\in V_1$ and $\cI(u)=\{\emptyset\}$ for $u\in V_2$.
\begin{remark}\label{r3}
An SP game is defined above as a {\em simultaneous-move} game: at each vertex $u\in V\setminus\{t\}$, players $1$ and $2$ choose strategies $\gs_1(u)\in\cI(u)$ and $\gs_2\in\cD(u)$ simultaneously. In an {\em alternating-move} version, player $1$ chooses first strategies $\gs_1(u)\in\cI(u)$, at each $u\in V\setminus\{t\}$, then player $2$ chooses an arc $(u,v)\in E(u)\setminus\gs_1(u)$, resulting in a play (which is either an $(s,t)$-path or an $s$-lasso). As we shall see later (proof of Corollary~\ref{c1}), the two versions are in fact equivalent, and a NE $(\gs_1,\gs_2)$ with respect to (wrt) the first definition corresponds to a NE $(\gs_1,\gs_2)$ wrt the second definition, which yields a play in the graph $G(\gs_1,\gs_2)$ defined by choosing an arbitrary path in $P_1(\sigma_1,\sigma_2)\cap P_2(\sigma_1,\sigma_2)$.
\end{remark}
\begin{remark}\label{r2}
Even though the definition of an SP interdiction game seems to be non-symmetric with respect to the two players, it can be in fact described in a symmetric way by defining the strategies of player 1 in terms of the ``dual'' independence system defined by its family of dependent sets $\cI^*(u)=\{E(u)\setminus I~|~I\in \cI(u)\}$, for $u\in V$. In this case, given a pair of strategies $(\sigma_1,\sigma_2)$, the play is defined by an $(s,t)$-path (or an $s$-lasso) in the subgraph defined by the arcs $\bigcup_{u\in  V\setminus\{t\}}\sigma_1(u)\cap\sigma_2(u)$. Equivalently, we may also define the game by the family of dependent sets $\cI^*$ of player 1 and the family of independent sets $\cD^*$ of player 2. 
\end{remark}

In terms of algorithms, since the number of independent sets at each vertex can be exponential, we will need to specify how the independence systems can be succinctly described. Assume that, for each vertex $u\in V\setminus\{t\}$, the family of independent sets $\cI(u)$ is given by an {\it independence oracle}, that is, an algorithm that decides, for any given $X\subseteq E(u)$, whether $X$ is independent (equivalently, $X\in\cI(u)$). For instance, we one can imagine a situation where the blocking of each arc $e\in E$ incurs some cost $c(e)>0$, and there is a budget of $\delta(u)$ at each vertex $u\in V\setminus\{t\}$ such that the blocking cost at $u$ cannot exceed $\delta(u).$ In such a case, the independence system at $u$ can be described by a linear inequality: $\cI(u)=\big\{I\subseteq E(u)~|~\sum_{e\in I}c(e)\le\delta(u)\big\}$.

\medskip

Theorem~\ref{t-Bi-SP-game} admits the following generalization.

\begin{theorem}\label{t-SP-interdiction}
Every finite positive two-person SP interdiction game has a NE (in pure stationary strategies). Assuming that the independence system at each vertex is given by a polynomial time oracle, such a NE can be computed in polynomial time. 
\end{theorem}


\subsection{Potential Transformations}
Out proofs rely heavily on the use of potential transformations defined as follows.
Given a graph $G=(V,E)$, a length function $r:E\mapsto \RR$ and a {\it potential} function
$\phi: V\mapsto \RR$, the potential transformation of $r$ is defined by
\[
	\br(u,v)=r(u,v)+\phi(v)-\phi(u) ~\text{ for all }~ (u,v)\in E.
\]
It is important to note that such potential transformation does not change the $r$-costs of the cycles, or the relative order of $(s,t)$-paths with respect to the total $r$-cost and fixed vertices $s$ and $t$. In particular, applying such transformations to the costs of the two players in an SP game does not affect the existence of a NE in the game. 

\paragraph{Notation.} For of a directed path $p$ in $G$ denote its vertex set by $V(p)$, and for vertices $u,w\in V(p)$, denote by $p[u,w]$ the subpath of $p$ starting at $u\in V(p)$ and ending at $w\in V(p)$.

\medskip

The rest of the paper contains the proofs of Theorems~\ref{t-Bi-SP-game},~\ref{t-Bi-SP-game-infinite} and~\ref{t-SP-interdiction}, in Sections~\ref{s-t-Bi-SP-game},~\ref{t-Bi-SP-game-infinite} and \ref{s-t-SP-interdiction}, respectively.
\section{Proof of Theorem~\ref{t-Bi-SP-game}}
\label{s-t-Bi-SP-game}

Consider an SP game $\Gamma=(G=(V_1\cup V_2,E),s,t,r_1,r_2)$.
Recall that $V_i$ is the set of positions controlled by player $i\in \{1,2\}$,  $s\in V_1\cup V_2$ is a fixed initial position, and $t\not\in V_1\cup V_2$ is a terminal position. Let us denote by $E_1$, $E_2$ the sets of possible moves of the players 1 and 2, respectively. Thus, $E_1\cup E_2=E$ and, by assumption (v),  $E_1\subseteq V_1\times (V_2\cup\{t\})$ and $E_2\subseteq V_2\times (V_1\cup\{t\})$. Let $r_i:E\mapsto \RR$ be the length function of the moves for player $i\in I$, and recall the following main assumptions:

\begin{itemize}
	\item[(A1)] For all positions $u\in V_1\cup V_2$ there exists at least one $(u,t)$-path (by assumption (iii)).
	\item[(A2)] The $r_i$-length of any directed cycle is positive for both players $i\in\{1,2\}$.
\end{itemize}

Recall that a strategy for player $i\in \{1,2\}$ is a mapping $\gs_i: V_i\to V_{3-i}$ 
such that for all $u\in V_i$ we have $(u,\gs_i(u))\in E_i$. A situation is a pair $(\gs_1,\gs_2)$ of strategies of the players. In every situation there is a unique walk starting at $s$, called the play of the situation and denoted by $P(\gs_1,\gs_2)$. This play may be a finite path terminating in $t$, in which case the cost of player $i\in I$ is the $r_i$-length of this path, or it maybe infinite (that is terminating in a directed cycle), in which case the cost for both players are $+\infty$, due to assumption (A2).

\begin{lemma}\label{t-1}
	Assume assumptions (A1) and (A2) and that the length functions satisfy the following conditions for all $i\in I$:
	\begin{itemize}
		\item[(L1)] For all $u\in V_i$ we have $0=\min\{r_i(u,v)\mid (u,v)\in E_i\}$.
		\item[(L2)] For all $u\in V_i$ there exists $(u,v)\in E_i$ s.t. $r_i(u,v)=0$ and $r_{3-i}(u,v)\leq 0$.
		\item[(L3)] For all $u\in V_i$ there exists $(u,v)\in E_i$ s.t. $r_{3-i}(u,v)=0$.
	\end{itemize}
	Then $\Gamma$ has a terminal NE.
\end{lemma}

\proof
First, for all $i\in \{1,2\}$ and for all $u\in V_i$ we choose an outgoing arc $(u,v)\in E_i$ such that $r_i(u,v)=0$ and $r_{3-i}(u,v)\leq 0$. By Condition (L2) we can choose such arcs. Denote by $H$ the subgraph formed by these arcs. 
Consequently, $H$ does not contain a cycle, since by assumption (A2) all cycles have positive $r_i$-length (for both $i\in \{1,2\}$). Therefore, there exists a directed $(s,t)$-path in $H$. Let us denote this path by $p$. 

Next, we define strategies for the players, as follows. For $i\in I$ and $u\in V_i\cap V(p)$ we define
$\gs_i(u)=v$ where $(u,v)$ is the arc of $p$ leaving $u$. For $u\in V_i\setminus V(p)$ we define $\gs_i(u)=v$ where $r_{3-i}(u,v)=0$. By condition (L3) we can choose such a $v$. 

Finally, we claim that the situation $(\gs_1,\gs_2)$ forms a terminal NE in $\Gamma$. Note that $p$ is the play of this situation. Since the role of the players is symmetric, assume for a contradiction that player $1$ can deviate and improve. Let us say that player $1$ deviates from $p$ and creates a new play $p'$ such that the $r_1$-length of $p'$ is strictly smaller than the $r_1$-length of $p$. Since both $p$ and $p'$ are simple paths, every arc can appear at most once in each of these paths. 

By our definition of $p$, the $r_1$-length of any arc in $p$ is nonpositive, and hence the total $r_1$-length of the arcs of $p$ that are not arcs of $p'$ is nonpositive. 

On the other hand, we have $r_1(u,V)\geq 0$ for all $u\in V_1$ and $(u,v)\in E_1$ by condition (L1), and have $r_1(u,\gs_2(u))=0$ for all $u\in V_2\setminus p$ by our definition of $\gs_2$. Thus the total $r_1$-length of arcs of $p'$ that are not arcs of $p$ is nonnegative, contradicting our assumption that $p'$ is a strict improvement over $p$. 
\qed

\medskip

We say that player $i\in\{1,2\}$ can block, if it has a strategy $\gs_i$ such that in the subgraph $\{(u,\gs_i(u))\mid u\in V_i\}\cup E_{3-i}$ there is a nonempty set $B\subseteq (V_1\cup V_{2})\setminus\{s\}$ with the property that no $(u,t)$-path exists for all $u\in B$. We say that player $i\in I$ can force $+\infty$, if it has a strategy $\gs_i$ such that in the subgraph $\{(u,\gs_i(u))\mid u\in V_i\}\cup E_{3-i}$ there is no $(s,t)$-path. Note that a player that cannot force $+\infty$ still may be able to block.

The next result is the simplest case when we can show the existence of a NE, under the strong condition that none of the players can block.

\begin{lemma}\label{t-2}
	Assume (A1), (A2), and that none of the players can block. 
 Then there exist (polynomially computable) potential functions 
 $\phi_i$, $i \in \{1,2\}$ such that the potential transformations
\[
\br_i(u,v)=r_i(u,v)+\phi_i(v)-\phi_i(u) ~\text{ for all }~ (u,v)\in E_1\cup E_2.
\]
satisfy conditions (L1), (L2), and (L3) for both $i\in\{1,2\}$. 
\end{lemma}

\proof 
By assumption, for both $i\in\{1,2\}$ and for all strategies of player $3-i$, the $r_i$-shortest paths have finite length for all positions. For a position $u\in V_1\cup V_2\cup \{t\}$ let us denote by $\phi_i(u)$ the length of the longest such $r_i$-shortest path, where we maximize over all strategies of player $3-i$. 
In fact (see~\cite{KGZ06,KBBEGRZ08}), 
for each $i \in\{1,2\}$  there exists a single tree rooted at $t$ that represents all such shortest-longest paths, and it can be computed in polynomial time. Note that $\phi_i$ satisfies the following relation for all positions:
\[
\phi_i(u) = 
\begin{cases}
	\min_{(u,v)\in E_i} r_i(u,v)+\phi_i(v)  \text{ if } u\in V_i,\\
	\max_{(u,v)\in E_{3-i}} r_i(u,v)+\phi_i(v)  \text{ if } u\in V_{3-i}.
\end{cases}
\]
It follows that for the transformed lengths we have for both $i\in\{1,2\}$ that
\begin{equation}\label{e-1}
	\begin{array}{rl}
		0&=\min \br_i(u,v) \text{ for } u\in V_i,\\
		0&=\max \br_i(u,v) \text{ for } u\in V_{3-i}.
	\end{array}
\end{equation}
Finally we note that (L1), (L2) and (L3) follow from \eqref{e-1}.
\qed

\begin{lemma}\label{t-3}
		Assume (A2) and that one of the payers cannot force $+\infty$. 
		Then $\Gamma$ has a terminal NE, which can be computed in polynomial time.
\end{lemma}

\proof
Wlog, we may assume that player $2$ cannot force $+\infty$. 

First we consider the $r_1$-shortest-longest paths, where we minimize over player $1$'s strategies, and maximize over strategies of player $2$. Let us denote by $\phi_1(u)$ the length of such an $r_1$-shortest-longest path from position $u\in V_1\cup V_2\cup\{t\}$. By \cite{KGZ06,KBBEGRZ08} we can determine in polynomial time
a directed tree rooted at $t$ that represents $r_1$-shortest-longest paths for every position for which $\phi_1$ is finite. Let us denote by
\[
B=\{u\in V_1\cup V_2\mid \phi_1(u)=+\infty\}
\]
the subset of the positions where $\phi_1$ is not finite,
and let $U=(V_1\cup V_2\cup \{t\})\setminus B$. Note that the set $B$ can also be computed in polynomial time (see \cite{KBBEGRZ08}). We denote by $E_1'\subseteq E_1$ and $E_2'\subseteq E_2$ the arc sets in the subgraph, induced by $U$. 

Then we have
\begin{equation}\label{e-phi1}
\phi_1(u) = 
\begin{cases}
	\displaystyle \min_{v : (u,v)\in E_1'} r_1(u,v)+\phi_1(v)  \text{ if } u\in V_1\setminus B,\\
	\displaystyle \max_{v : (u,v)\in E_2'} r_1(u,v)+\phi_1(v)  \text{ if } u\in V_2\setminus B.
\end{cases}
\end{equation}

Note that by our assumptions, we have $s\in U$. Furthermore, player $1$ cannot leave the set $B$, while player $2$ cannot enter it (from $U$). Thus, we have the following properties:
\begin{equation}\label{e-Bprop}
	\begin{array}{l}
		\forall u\in V_1\cap B, ~ (u,v)\in E_1 \text{ we have } v\in B,\\
		\nexists u\in V_2\cap U, ~ (u,v)\in E_2 \text{ with } v\in B,\\
		\forall u\in V_2\cap B ~\exists (u,v)\in E_2 \text{ with } v\in B.
	\end{array}
\end{equation}

We then use $\phi_1$  and apply a potential transformation to the $r_1$ length function in the subgraph induced by $U$. 
Then for the transformed lengths we have by \eqref{e-phi1} that
\begin{subequations}\label{e-br1}
	\begin{align}
		0&=\displaystyle \min_{(u,v)\in E_1'} \br_1(u,v) \text{ for } u\in V_1\setminus B,\label{e-br1-a}\\
		0&=\displaystyle \max_{(u,v)\in E_2'} \br_1(u,v) \text{ for } u\in V_2\setminus B.\label{e-br1-b}
	\end{align}
\end{subequations}
Note that we have no precise information about $\br_1(u,v)$ for $(u,v)\notin E_1'\cup E_2'$.
Nevertheless, by equations \eqref{e-br1} we have for all $u\in (V_1\cup V_2)\setminus B$ a move $(u,v)\in E_1'\cup E_2'$ such that $\br_1(u,v)=0$. For $u\in V_1$, let us define
\begin{equation}
	\gs_1(u) = 
	\begin{cases}
		v & \text{ for } u\in V_1\setminus B \text{ such that } (u,v)\in E_1', ~ \br_1(u,v)=0,\\
		v & \text{ for } u\in B \text{ arbitrarily such that } (u,v)\in E_1.
	\end{cases}
\end{equation}
By \eqref{e-Bprop} and \eqref{e-br1-a} strategy $\gs_1$ is well-defined.
We introduce the notation 
\begin{equation}\label{e-H}
	H=\{(u,\gs_1(u))\mid u\in V_1\setminus B\} 
\end{equation}
for the set of $\gs_1$-moves of player $1$ within $U$. 

The subgraph $(U,H\cup E_2')$ now contains the spanning $r_1$-shortest-longest paths we constructed within $E_1'\cup E_2'$. Thus the 
$r_2$-shortest 
$(u,t)$-path in the subgraph $(U,H\cup E_2')$ is finite for each $u\in U$. 

Let us denote by $p$ an $r_2$-shortest $(s,t)$-path, and for $u\in V_2$, define 
\[
\gs_2(u)=
\begin{cases}
v & \text{ for } u\in V(p)\cap V_2  \text{ such that } (u,v)\text{ is an arc of } p ,\\
v & \text{ for } u\in V_2\setminus (B\cup V(p)) \text{ such that }	\br_1(u,v)=0 ,\\
v & \text{ for } u\in V_2\cap B	\text{ such that }v \in B,~ (u,v) \in E_2 . 	 
\end{cases}
\]
By \eqref{e-br1-b} we have a $v$ for all $u\in V_2\setminus V(p)$ such that $\br_1(u,v)=0$. By \eqref{e-Bprop} for every $u\in V_2\cap B$ we have a $(u,v)\in E_2$  such that $v\in B$. Thus, $\gs_2$ is a well-defined strategy of player $2$. 

We claim that the situation $(\gs_1,\gs_2)$ is a terminal NE. Note that $p$ is the play of this situation, and hence it is terminal. 

Let us first observe that, by \eqref{e-H}, player $2$ could improve on $p$ only if there exists an $(s,t)$-path within the subgraph $(U,H\cup E_2')$, which has a shorter $r_2$-length than $p$. This is however impossible, since $p$ was chosen as an $r_2$-shortest $(s,t)$-path.  Furthermore, by \eqref{e-Bprop} player $2$ cannot enter $B$. Thus player $2$ cannot change $\gs_2$ and improve strictly.

As for player $1$, let us note first that for all $u\in V(p)\cap V_1$ we have $\br_1(u,\gs_1(u))=0$ and for all $v\in V_2\cap V(p)$ we have $\br_1(v,\gs_2(v))\leq 0$ by our definitions of $\gs_1$ and $\gs_2$, and by \eqref{e-br1-b}. Thus for player $1$ every move along $p$ has a nonpositive $\br_1$-length. 

For his/her moves outside $p$, we have $\br_1(u,\gs_2(u))=0$ for all $u\in V_2\setminus(B\cup V(p))$ by our definition of $\gs_2$, and for all moves $(u,v)\in E_1'$, $u\in V_1\cap U$ we have $\br_1(u,v)\geq 0$ by \eqref{e-br1-a}. Thus all player $1$ can do is to replace some nonpositive moves along $p$ by some nonnegative moves outside $p$, and hence he/she cannot improve either. This proves our claim that $(\gs_1,\gs_2)$ is a NE, completing the proof of the lemma. 
\qed

As mentioned earlier, when both players can force $+\infty$, then the pair of forcing strategies constitutes a (cyclic) NE. As can be seen from the proof of Lemma~\ref{t-3}, such $+\infty$-forcing strategies for the two players can be found in polynomial time by computing the $r_i$-shortest-longest paths, and checking if the lengths obtained satisfy $\phi_i(s)=+\infty$, for $i\in\{1,2\}$. This concludes the proof of Theorem~\ref{t-Bi-SP-game}.

\begin{remark}\label{r1}
	Let us add that we could have used the lengths of the shortest $r_2$-paths in the second part of the above proof as a potential function, and transform with it the $r_2$-lengths of the arcs in $E_2'$. It could then be shown that in this situation properties (L1), (L2), and (L3) hold in the subgraph induced by $U$. Thus, there exists a terminal NE within this subgraph by Lemma \ref{t-1}. Using properties \eqref{e-Bprop} then we could have argued that this terminal NE can be extended to a NE in $\Gamma$. 
\end{remark}

\section{Proof of Theorem~\ref{t-Bi-SP-game-infinite}}\label{s-t-Bi-SP-game-infinite}

Due to compactness arguments, Theorem~\ref{t-Bi-SP-game} holds also in the case of infinite digraphs such that all vertices have a finite out-degree. In this case  infinite plays may be infinite paths with finite effective costs.

Let $G=(V,E)$ be an infinite digraph, $s$ be the initial vertex and $\Gamma=(G,s,t,r_1,r_2)$ be an SP game defined on $G$. We assume that $V$ is countable and fix an enumeration of vertices: $v_0, v_1,\dots$, where $v_0=s$. For each $n$ define a finite part of the game $\Gamma^n=(G^n,s,t,r_1,r_2)$ defined on the finite digraph $G^n=(V^n, E^n)$ with vertex set
\[
V^n =\bigcup_{i=0}^{n-1} \big(\{v_i\}\cup N^+(v_i)\big), \quad\text{where}\
N^+(v) = \{u: (v,u)\in E\}.
\]
and having arc set
\[
E^n = \{(v_j,v_r): 0\leq j<n, (v_j, v_r)\in E\}.
\]
Players control the same vertices in the set $\{v_0, \dots, v_{n-1}\}$ as in the infinite game $\Gamma$ and vertices $v_r$, $r\geq n$, are terminals in $\Gamma^n$ (and local costs are inherited from $\Gamma$).

By Theorem~\ref{t-Bi-SP}, there exists a NE for each $\Gamma^n$. Fix NE strategy profiles $(\ts^n_1, \ts^n_2)$ for each $\Gamma^n$. Construct a pair $(\ts^\infty_1, \ts^\infty_2)$ of strategies for infinite game 
such that for all $m$ and for infinitely many values of $n$ the restriction of $(\ts^\infty_1, \ts^\infty_2)$ on the
game $\Gamma^m$ coincides with the restriction of  $(\ts^n_1,\ts^n_2)$ on the game $\Gamma^m$.
It can be done inductively. For each $m$ there are finitely many  strategy profiles for~$\Gamma^m$. Hence it is possible to choose $(\gs^m_1,\gs^m_2)$ such that $(\gs^m_1,\gs^m_2)$  is the restriction of $(\ts^n_1, \ts^n_2)$ for infinitely many~$n$. Once  $(\gs^m_1,\gs^m_2)$ is chosen, we continue for $m'>m$ choosing among strategy profiles whose restrictions on $\Gamma^m$ coincide with $(\gs^m_1,\gs^m_2)$.

We claim that $(\ts^\infty_1, \ts^\infty_2)$ is a NE in
$\Gamma$. Suppose for the sake of contradiction that some player~$i\in\{1,2\}$ can
deviate and improve. For a path $p$ in the game $\Gamma$ we denote by
$p[m]$ the initial part of $p$ inside the game $\Gamma^m$ (it finishes
when $p$ enters first time a vertex $v_j$, $j\geq m$).
Let $p$ be the deviated path; $\tp$ is the
path generated by $(\ts^\infty_1, \ts^\infty_2)$; and $\tp^n$ be the
path generated by $(\ts^n_1, \ts^n_2)$.

By construction of $(\ts^\infty_1, \ts^\infty_2)$, for any $t$, the
path $\tp[t]$ is the initial part of the path $\tp^n$ for infinitely
many $n$. Thus $R_i(\tp^n)\geq R_i(\tp[t])$ for infinitely
many $n$.  Note that $R_i(p[n])\geq R_i(\tp^n)$, since $(\ts^n_1,\ts^n_2)$ is a NE in the game
$\Gamma^n$. Therefore, for all $t$ and infinitely many $n$,
$R_i(p[n])\geq R_i(\tp[t])$.
Since $R_i(p) = \lim_{n\to\infty}R_i[n]$ and $R_i(\tp) = \lim_{t\to\infty} R_i(\tp[t]) $, we  get $R_i(p)\geq R_i(\tp)$, which contradicts to the assumption that player $i$ improves. 


\section{Proof of Theorem~\ref{t-SP-interdiction}}\label{s-t-SP-interdiction}
\subsection{Existence of NE}
The following result follows from Theorem~\ref{t-Bi-SP-game}.
\begin{corollary}\label{c1}
Every finite positive two-person SP interdiction game has a NE (in pure stationary strategies). 
\end{corollary}
\proof
We reduce any given SP interdiction game $\Gamma=(G=(V,E),\cI,r_1,r_2)$
to an SP game $\Gamma'=(G'=(V'=V_1'\cup V_2',E'),r_1',r_2')$ as follows (see~\cite{Pis99} for a similar reduction in the case of zero-sum mean payoff game). For every vertex $u\in V$, and every independent set $I\in\cI(u)$, we introduce a new vertex $u_I\in V'$ and an arc $(u,u_I)\in E'$. We also define an arc $(u_I,v)\in E'$ for every pair of vertices $u,v\in V$ and independent set $I\in\cI(u)$ such that $(u,v)\in E(u)\setminus I(u)$: 
\begin{align*}
    V_1'&=V,\ \ \ V_2'=\{u_I:~u\in V,~I\in\cI(u)\},\ \ \ V'=V_1'\cup V_2',\\
    E'&=\{(u,u_I):~v\in V\setminus\{t\},~I\in\cI(v)\}\cup\{(u_I,v):~u\in V\setminus\{t\},~I\in\cI(u),~(u,v)\in E(u)\setminus I\}.
\end{align*}
For $i\in\{1,2\}$, we set $r_i'(v,v_I)=0$ for $v\in V\setminus\{t\},~I\in\cI(v)$, and $r_i'(u_I,v)=r_i(u,v)$ for $(u,v)\in E(u)\setminus I,~I\in\cI(u)$.

We make the following simple observations:
\begin{itemize}
    \item [(o1)] Any $(s,t)$-path (possibly an $s$-lasso) in $G'$ corresponds to an $(s,t)$-path in $G$ with the same $r_i$-cost, for $i\in\{1,2\}$.

 
    \item [(o2)] For every pair of strategies $(\sigma_1,\sigma_2)$ in $\Gamma$, we can define a pair of strategies $(\sigma_1',\sigma_2')$ in $\Gamma'$ as follows: for  $u\in V\setminus\{t\}$ define $\sigma_1(u)=u_{\sigma_1(u)}$, and for $I\in\cI(u)$, define $\sigma_2'(u_{I})=v$ for an arbitrary arc $(u,v)\in\sigma_2(u)\setminus I$.  

     \item [(o3)] Given a pair of strategies $(\sigma_1',\sigma_2')$ in $\Gamma'$, we can define a pair of strategies $(\sigma_1,\sigma_2)$ in $\Gamma$ as follows: for $u\in V\setminus\{t\}$, where $\sigma_1'(u)=u_{I}$, define $\sigma_1(u)=I$ and $\sigma_2(u)=\bigcup_{I'\in\cI(u)}\{(u,\sigma_2'(u_{I'})\}$. By construction of $G'$, $\sigma_2(u)\in\cD(u)$ since $\sigma_2(u)\setminus I'\neq\emptyset$ for all $I'\in\cI(u)$ and $u\in V\setminus\{t\}$. 
    
    \item [(o4)] Let $(\sigma_1',\sigma_2')$ be a pair of strategies which forms a NE in $\Gamma'$. Consider the corresponding pair of strategies $(\sigma_1,\sigma_2)$ defined in $\Gamma$ as described in (o3).  
We claim that the pair $(\sigma_1,\sigma_2)$ forms a NE in $\Gamma$. Clearly, player $2$ cannot deviate in $\Gamma$ and improve by selecting another strategy $\widehat\sigma_2$, since this would imply that the corresponding strategy $\widehat\sigma_2'$, defined in (o2), yields a smaller $r_2$-cost for player $2$ in $\Gamma'$, implying that $\widehat\sigma_2'$ is an improving strategy for player $2$ in $\Gamma'$. Suppose player 1 can improve by deviating to another strategy $\widehat\sigma_1$ in $\Gamma$. Consider the improving $(s,t$)-path $p$ in $G$ and let $p'$ be the corresponding path in $G'$ as described in (o1). It is possible that the path $p'$ uses an arc $(u_{\widehat I},v)$ for some $u,v\in V$ such that $\widehat I=\widehat\sigma_1(u)\in\cI(u)$ and $v\ne\sigma_2'(u_{\widehat I})$. But since $(u,v)\in \sigma_2(u)\setminus\widehat\sigma_1(u)$, there must exist another $I\in\cI(u)$ such that $\sigma_2'(u_I)=v$. Redefining $\widehat\sigma_1(u)=I$ does not change the cost of the path $p'$, and doing a similar update for every vertex $u$ on the path $p$ ensures that the modified path $p'$ in $G'$ uses only arcs defined by $\sigma_2'$. Thus if player $1$ switches to the modified strategy $\widehat\sigma_1$ obtained after this sequence of updates, he ensures  a strategy $\widehat\sigma_1'$ in $\Gamma'$ that leads to a smaller $r_1$-cost $(s,t)$-path, in response to $\sigma_2'$, in contradiction to the assumption that  $(\sigma_1',\sigma_2')$ is a NE in $\Gamma'.$ 
    \item [(o5)] If $G$ has positive local costs, then all directed cycles in $G'$ are also positive (as all such cycles must include at least two arcs of the form $(u_I,v)$, for some $u\in V\setminus\{t\}$, $I\in \cI(u)$ and $v\in E(u)\setminus\cI(u)$, which have positive costs), implying by the result in~\cite{Gal58} that all local costs in $G'$ can be made positive by a potential transformation. 
\end{itemize}
The corollary follows from the above observations and Theorem~\ref{t-Bi-SP-game}. 
\qed

Note that the reduction used in the proof of Corollary~\ref{c1} does not yield a polynomial-time algorithm for computing a NE in the interdiction game, since the number of independent sets at any vertex can be exponential.

\subsection{A Polynomial-time Algorithm}
Consider an SP interdiction game $\Gamma=(G=(V,E),s,t,\cI,r_1,r_2)$.  Assume that, for each vertex $u\in V\setminus\{t\}$, the family of independent sets $\cI(u)$ is given by an independence oracle and denote by $\tau$ the maximum time taken by the oracle on any set $X\subseteq E(u)$, over all $u\in V$. Any subgraph of $G$ obtained by removing a subset of arcs $I(u)\in\cI(u)$, at every $u\in V$, will be called {\it admissible}. We recall the following result, implicitly proved in~\cite{KGZ06,KBBEGRZ08}.

\begin{lemma}\label{l1}
Given a directed graph $G=(V,E)$ with associated non-negative arc lengths $r:E\to \RR_+$, and a target vertex $t$, there is an $O(|E|(\log|V|+\tau))$ algorithm (Modified Dijkstra's) that computes a potential function $\phi:V\to\RR$, and independent sets $\widehat I(u)\in\cI(u)$, for $u\in V\setminus\{t\}$, such that 
\begin{subequations}\label{ei-br1}
\begin{align}
\phi(u)&\ge\displaystyle r(u,v)+\phi(v)\text{ for all  $(u,v)\in\widehat I(u)$,}\label{ei-br1-a}\\
\phi(u)&\le r(u,v)+\phi(v),\text{ for all $(u,v)\in E(u)\setminus\widehat I(u)$,}\label{ei-br1-b}\\
\phi(u)&=\displaystyle r(u,v_u)+\phi(v_u),\text{ for some $(u,v_u)\in E(u)\setminus\widehat I(u)$,}\label{ei-br1-c}\\
\cI(u)&\not\ni\{(u,v)\in E(u)~|~\phi(u)\ge r(u,v)+\phi(v)\}.\label{ei-br1-d}
\end{align}
\end{subequations}
\end{lemma}
Note that conditions (\ref{ei-br1-b})-(\ref{ei-br1-d}) imply that 
$\phi(u)$ is the maximum shortest $r$-distance from $u$ to $t$ in any admissible subgraph of $G$. Note also that there is {\it one} admissible graph which maximizes all such distances from all vertices to $t$.
For completeness, 
we include a proof of Lemma~\ref{l1} in the appendix.

We say that player $1$ can force $+\infty$ in an SP interdiction game, if it has a blocking strategy that prevents player $2$ from reaching the terminal starting from $s$. A terminal NE is a NE in which the costs of both players are finite.

\begin{lemma}\label{l2}
Consider an SP interdiction game $\Gamma=(G=(V,E),s,t,\cI,r_1,r_2)$, in which player $1$ cannot force $+\infty$. Then $\Gamma$ has a terminal NE that can be computed in $O(|E|\log|V|+\tau|E|)$ time.
\end{lemma}
\proof
Let $\phi_2$ be the potential function computed in Lemma~\ref{l1} with respect to the cost function $r=r_2$, and let $\widehat I(u)$, for $U\in V\setminus\{t\}$, be the obtained blocking sets. Let $B=\{u\in  V~|~\phi_2(u)=+\infty\}$, $U=V\setminus B$, and denote by $E'\subseteq E$ the arc set in the subgraph induced by $U$. 

Note that by our assumption, we have $s\in U$. Furthermore, player $1$ has a strategy that prevents player $2$ from leaving the set $B$, but none that forces entering it (from $U$). Thus, we have the following properties:
\begin{equation}\label{ei-Bprop}
	\begin{array}{l}
		\forall u\in B, ~ \{(u,v)\in E(u)~|~v\in U\}\in \cI(u),\\
		\forall u\in U, ~ \{(u,v)\in E(u)~|~v\in U\}\not\in \cI(u).
	\end{array}
\end{equation}
We then use $\phi_2$ and apply a potential transformation to the 
$r_2$-length 
function in the subgraph induced by $U$. 
Then for the transformed lengths $\br_2$ we have by Lemma~\ref{l1} that, for all $u\in U\setminus\{t\}$,
\begin{subequations}\label{ei-br2}
\begin{align}
\br_2(u,v)&\le 0\displaystyle \text{ for all  $(u,v)\in\widehat I(u)$,}\label{ei-br2-a}\\
\br_2(u,v)&\ge 0,\text{ for all $(u,v)\in E(u)\cap E'\setminus\widehat I(u)$,}\label{ei-br2-b}
\\
\br_2(u,v_u)&=0\displaystyle,\text{ for some $(u,v_u)\in E(u)\cap E'\setminus\widehat I(u)$,}\label{ei-br2-c}\\
\cI(u)&\not\ni\{(u,v)\in E(u)\cap E'~|~\br_2(u,v)\le 0\}.\label{ei-br2-d}
\end{align}
\end{subequations}
Let $E''=\{(u,v)\in E'~|~\br_2(u,v)\le 0\}$ and note that~\raf{ei-br2-c} (and the absence of $0$-cycles in $G$) implies that there is an 
$(s,t)$-path 
in $G'=(U,E'')$.
With every $(u,v)\in E(u)\cap E''$, let us associate an independent set $I(u,v)=\{(u,v')\in\widehat I(u)~|~\br_2(u,v')<\br_2(u,v)\}$, and note by this definition 
that 
\begin{align}\label{l1-e3}\br_2(u,v)&=\min_{(u,v')\in E(u)\cap E'\setminus I(u,v)}\br_2(u,v').
\end{align}

Next, let $p$ be an $r_1$-shortest $(s,t)$-path in the graph $G'$, and  for $u\in V$, define 
\[
\gs_1(u)=
\begin{cases}
I(u,v), & \text{ for } u\in V(p)  \text{ such that } (u,v)\text{ is an arc of } p ,\\
\widehat I(u), & \text{ for } u\in U\setminus  V(p),\\
\{(u,v)\in E(u)~|~v\in U\},  & \text{ for } u\in B. 	 
\end{cases}
\]
Note that (\ref{ei-Bprop}) implies that $\sigma_1$ is well-defined. For $u\in V\setminus\{t\}$, we define  
\[
\sigma_2(u)=
\begin{cases}
E(u)\cap E'', & \text{ for } u\in U,\\
D\in\cD(u),  & \text{ for } u\in B. 	 
\end{cases}
\]
Note by~\raf{ei-br2-d} that $\sigma_2(u)\in\cD(u)$.  
We claim that the situation $(\gs_1,\gs_2)$ is a terminal NE. Note that $p$ is an $(s,t)$-path in the graph $G(\sigma_1,\sigma_2)$, and hence to show that the situation $(\sigma_1,\sigma_2)$ is a terminal NE, it remains to prove that $p$ is an $r_i$-shortest path in the game $G(\sigma_1,\sigma_2)$. 

Observe next that, by the definition of $\sigma_2$ and~\eqref{ei-Bprop}, player $1$ cannot force entering $B$. It follows then, by the definition of $\sigma_1$ and $\sigma_2$ that player $1$ could improve on $\sigma_1$ only if there exists an $(s,t)$-path within the subgraph $G'$, which has a shorter $r_1$-length than $p$. This is however impossible, since $\sigma_1$ was chosen such that the subgrapgh $G(\sigma_1,\sigma_2)$ contains an $r_1$-shortest $(s,t)$-path. Thus player $1$ cannot change $\gs_1$ and improve strictly.

As for player $2$, suppose he can improve strictly by deviating to a strategy $\sigma_2'$ such that there is a path $p'$ in the graph $G(\sigma_1,\sigma_2')$ with strictly less $r_2$-cost than $p$. Note that player $2$ is not interested in entering $B$ as he cannot escape from there given the choice of $\sigma_1$. Thus all the arcs in $p'$ should lie inside $U$. 
Consider any arc $(u,v)\in p\setminus p'$ and let  $(u,v')$ be the arc at $u$ that belongs to $p'$.
 The definition of $\sigma_1$ and~\raf{l1-e3} imply that $\br_2(u,v')\ge\br_2(u,v)$.  

For arcs $(u,v)$ of $p'$ such that $u\not\in V(p)$, we have $\br_2(u,v)\geq 0$ by \eqref{ei-br2-b}. Then, by the above observations and the non-positivity of the arcs on $p$,
\begin{align*}
\br_2(p'\setminus p)&=\sum_{(u,v)\in p'\setminus p:~u\in V(p)}\br_2(u,v)+\sum_{(u,v)\in p'\setminus p:~u\not\in V(p)}\br_2(u,v)\\
&\ge\sum_{(u,v)\in p\setminus p':~u\in V(p')}\br_2(u,v)+\sum_{(u,v)\in p\setminus p':~u\not\in V(p')}\br_2(u,v)=\br_2(p\setminus p').
\end{align*}

Thus all player $2$ can do is to replace some moves of $p$ by some equal or costlier moves outside $p$, and hence he/she cannot improve either. This proves our claim that $(\gs_1,\gs_2)$ is a NE, completing the proof of the lemma.
\qed

\paragraph{Proof of Theorem~\ref{t-SP-interdiction}.} Consider an SP interdiction game $\Gamma=(G=(V,E),s,t,\cI,r_1,r_2)$. If both players can force $+\infty$, then the corresponding strategies define a (cyclic) NE. Such strategies can be computed in $O(|E|\log|V|+\tau|E|)$ by applying Modified Dijkstra's algorithm as in Lemma~\ref{l1}. If player 1 cannot force $+\infty$, then a terminal NE exists by Lemma~\ref{l2}. If player 2 cannot force $+\infty$, then in view of Remark~\ref{r2}, we can consider the SP interdiction game defined by the dual independence system and conclude again by Lemma~\ref{l2} that it has a terminal NE. The theorem follows.  

\section*{Acknowledgments} 
The third and forth authors were working 
within the framework of the HSE University Basic Research Program. 
The fourth author was supported in part by the state assignment topic no. FFNG-2024-0003.

\appendix
\section{Proof of Lemma~\ref{l1}}
The algorithmic proof is based on the following modification of Dijkstra's algorithm from~\cite{KGZ06,KBBEGRZ08}.

\RestyleAlgo{ruled}
\begin{algorithm}

\caption{Modified Dijkstra's Algorithm $(G=(V,E),t,\cI,r)$}
	\KwData{A digraph $G=(V,E)$ with arc weights $r:E\to\RR_+$, a terminal vertex $t\in V$, and a blocking system $\cI\in 2^{E}$ defined by an independence oracle. }
	\KwResult{A potential function $\phi:V\to\RR$, and independent sets $\{\widehat I(u)\in\cI(u)~|~u\in V\setminus\{t\}\}$, 
    satisfying~\eqref{ei-br1}}
// {\it Initialization}:\\
$S=V\setminus\{t\}$; $T=\{t\}$\\
$\phi(u)=+\infty$ for $u\in S$; $\phi(t)=0$\\
$\widehat I(u)=\emptyset$ for $u\in S$\\
Initialize an empty binary heap $\cH$\\
\For{each arc $e=(v,t)\in E$}{Insert $e$ into $\cH$ with key $k(e)=r(e)$}

// {\it Iteration loop}:\\
\While{$\cH\ne\emptyset$}{
Extract the minimum-key arc $e=(u,v)$ from $\cH$\label{s11}\\
\If{$u\in S$ and $v\in T$}{
\eIf{$\widehat I(u)\cup\{e\}\in\cI(u)$\label{s13}}{
$\widehat I(u)=\widehat I(u)\cup\{e\}$\label{s14}
}
{ $S=S\setminus\{u\}$; $T=T\cup\{u\}$;  $\phi(u)=k(e)$\label{s16}\\
\For{each arc $e=(v,u)\in E$ such that $v\in S$}{
Insert $e$ with key value $k(e)=r(e)+\phi(u)$ into $\cH$
\label{s18}}
}
} 
}
\Return $\phi$,  $\{\widehat I(u)~|~u\in V\setminus \{t\}\}$
\end{algorithm}

It easy to see that the running time of the algorithm is $O(|E|(\log|V|+\tau))$. To see that the function $\phi$, and independent sets $\{\widehat I(u)\in\cI(u)~|~u\in V\setminus\{t\}\}$ output by the algorithm satisfy~\raf{ei-br1}, let us show that the algorithm maintains the invariant that~\raf{ei-br1} hold for all $u\in T\setminus\{t\}$. Initially, when $T=\{t\}$, the invariant holds trivially.  For the sake of a contradiction, consider the first iteration of the algorithm in which the invariant is violated. Let $e=(u,v)$ be the extracted arc in step~\ref{s11} in that iteration. Clearly, the invariant remains valid if $u$ is not moved from $S$ to $T$. Assume therefore, that steps~\ref{s16}-\ref{s18} are executed and let us denote in this case $e=(u,v_u)$. Then, by the updates used to construct $\widehat I(u)$ in step~\ref{s14} and the condition checked in step~\ref{s13}, we get that
\begin{align}\label{l1-e1}
    \widehat I(u)\cup\{(u,v_u)\}\not\in \cI(u).
\end{align}
By the update in step~\ref{s16} and the fact that arcs across the $(S,T)$-cut are extracted from $H$ in increasing key-order, we have 
\begin{align}\label{l1-e2}
    r(u,v_u)+\phi(v_u)=\phi(u)\ge r(u,v)+\phi(v),\text{ for all }(u,v)\in\widehat I(u).
\end{align}
Thus,~\raf{ei-br1-a},~\raf{ei-br1-c} and~\raf{ei-br1-d} hold by~\raf{l1-e1} and~\raf{l1-e2}. Now consider~\raf{ei-br1-b}. For any arc $(u,v)\in E(u)\setminus\widehat I(u)$ with $v\in S$, we have $\phi(v)=+\infty$, and therefore $r(u,v)+\phi(v)\ge\phi(u)$ holds trivially. On the other hand, any arc $(u,v)\in E(u)\setminus\widehat I(u)$ with $v\in T$ was considered among the arcs crossing the $(S,T)$-cut, prior to moving $u$ to $T$, but was not extracted from $\cH$. This implies that $k(u,v)=r(u,v)+\phi(v)\ge k(e)=\phi(u)$. We conclude that~\raf{ei-br1-b} holds for $u$. Let us next consider a vertex $u'\in T$ such that $(u',u)\in E(u')$ and suppose that after updating $\phi(u)$, we get $\phi(u')>r(u',u)+\phi(u)$. The fact that~\raf{ei-br1-c} holds for all $v\in T\cup\{u\}\setminus\{t\}$ implies that there is a $(u,t)$-path $p=\{(u_0,u_1),\ldots,(u_{k-1},u_k)\}$ in $G$ such that $u_0=u$, $u_k=t$ and $\phi(u_{i-1})=r(u_{i-1},u_{i})+\phi(u_{i})$, for $i=1,\ldots,k$. Consider the moment when $u'$ was moved from $S$ to $T$; there existed a vertex on $p$, say $u_j$, such that the arc $(u_j,u_{j+1})$ belonged to the $(S,T)$-cut at that moment, but was not extracted from $\cH$. This implies that $\phi(u')\le r(u_j,u_{j+1})+\phi(u_{j+1})$. We get
\[
r(u',u)+r(p[u,u_{j+1}])+\phi(u_{j+1})=r(u',u)+\phi(u)<\phi(u')\le r(u_j,u_{j+1})+\phi(u_{j+1})
\]
which contradicts the non-negativity of $r$.
Finally, note that when $\cH$ becomes empty, we have $\phi(u)=+\infty$ for every $u\in S$. Furthermore, all arcs $(u,v)\in E(u)$ with $u\in S$ and $v\in T$ are contained in $\widehat I(u)$. This together with the assumption that $E(u)\not\in \cI(u)$ imply that any $u\in S$ must have an arc $(u,v_u)\in S$. Thus, \raf{ei-br1} also holds for all $u\in S$ at the end of the algorithm. The lemma follows.

\end{document}